\documentstyle[12pt]{article}
\newcommand{\Ir}{Z\!\!\!Z}
\newcommand{\Ibb}[1]{ {\rm I\ifmmode\mkern
            -3.6mu\else\kern -.2em\fi#1}}
\newcommand{\ibb}[1]{\leavevmode\hbox{\kern.3em\vrule
     height 1.2ex depth -.3ex width .2pt\kern-.3em\rm#1}}
\newcommand{\Cx}{{\ibb C}}
\newcommand{\Ham}{{\Ibb H}}
\newcommand{\Rl}{{\Ibb R}}
\newcommand{\be}{\begin{eqnarray}}
\newcommand{\ee}{\end{eqnarray}}
\renewcommand{\O}{\Omega}
\newcommand{\A}{{\cal A}}
\newcommand{\B}{{\cal B}}
\newcommand{\I}{{\cal I}}
\renewcommand{\S}{{\bf S}}
\renewcommand{\d}{\mbox{d}}
\newcommand{\cst}{\mbox{\small const}}
\newcommand{\pa}{\partial}
\newcommand{\bu}{\bullet}
\newcommand{\oc}{\otimes}
\begin{document}

\begin{tabbing}
\hspace*{12cm}\= GOET-TP 100/95  \\
              \> December 1995
\end{tabbing}
\vskip.2cm

\centerline{\LARGE \bf Integrable Discretizations of Chiral Models}
\vskip.5cm

\centerline{\LARGE \bf via Deformation of the Differential Calculus}
\vskip1cm

\begin{center}
      {\large \bf Aristophanes Dimakis} \ and \ {\large \bf Folkert
       M\"uller-Hoissen}
       \vskip.3cm
      Institut f\"ur Theoretische Physik  \\
      Bunsenstr. 9, D-37073 G\"ottingen, Germany
\end{center}
\vskip.8cm

\begin{abstract}
\noindent
A construction of conservation laws for chiral models (generalized
$\sigma$-models) on a two-dimensional space-time continuum using
differential forms is extended in such a way that it also comprises
corresponding discrete versions. This is achieved via a deformation of
the ordinary differential calculus. In particular, the nonlinear Toda
lattice results in this way from the linear (continuum) wave equation.
The method is applied to several further examples. We also construct
Lax pairs and B\"acklund transformations for the class of models
considered in this work.
\end{abstract}

\section{Introduction}
Some years ago we observed that a certain deformation of the ordinary
calculus of differential forms on $\Rl^n$ can be used to discretize
classical continuum field theories \cite{DMHS93}. For this purpose one
has to formulate the theory in terms of differential forms. The
deformation of the differential calculus then induces a corresponding
deformation of the theory built on it.
In particular, the Wilson loop formulation of lattice gauge theory
originates in this way from continuum Yang-Mills theory. In this letter
we present another application of the method, a discretization of
chiral models\footnote{See \cite{Pere87} and the references cited
there.} (or generalized $\sigma$-models) preserving complete
integrability.
After a brief introduction to deformations of the ordinary differential
calculus on $\Rl^2$ we generalize the derivation of conservation laws
given for chiral models in \cite{Bre79}.
As an example, we then derive the nonlinear Toda lattice \cite{Toda}
from the continuum wave equation. Further examples illustrate the
method and reveal its limitations. We also construct Lax pairs
for the discretized chiral models and present B\"acklund
transformations.

\section{Deformation of the ordinary differential calculus on $\Rl^2$}
In the ordinary differential calculus on manifolds, functions commute
with differentials. It is possible, however, to dispense with this
property while keeping the familiar rules for the exterior derivative
(see \cite{DMHS93} for details).
The latter are consistent with the following commutation relations,
\be
      \d t \,  f(t,x) = f(t+\tau,x) \, \d t \; , \qquad
      \d x \,  f(t,x) = f(t,x+\lambda) \, \d x
\ee
where $\tau$ and $\lambda$ are real parameters, $f$ is a function
on $\Rl^2$ and $t,x$ are the canonical coordinate functions on $\Rl^2$.
In particular,
\be
     [ \d t,t ] = \tau \, \d t \; , \qquad [ \d t,x ] = [\d x,t] = 0
     \; , \qquad
     [ \d x,x ] = \lambda \, \d x \; .       \label{cr}
\ee
Using these relations we obtain for $\tau,\,\lambda \neq 0$
\be
      \d f  =  (\pa_{+t}f) \, \d t + (\pa_{+x}f) \, \d x
              =  \d t \, (\pa_{-t}f) + \d x \, (\pa_{-x} f)
\ee
with
\be
     \pa_{+x}f(t,x) = {1\over\lambda} [f(t,x+\lambda)-f(t,x)]
     \; , \qquad
     \pa_{-x}f(t,x) = {1\over\lambda} [f(t,x)-f(t,x-\lambda)]
\ee
and similar expressions for $\pa_{\pm t}f(t,x)$. This shows that for
$\tau,\,\lambda \neq 0$ the differential calculus is actually defined
on the algebra $\A$ of {\em all} real functions on $\Rl^2$.\footnote{In
order to perform the limit $\tau\to 0$ (or $\lambda \to 0$) we have to
restrict $\A$ to those functions which are differentiable in $t$
(respectively, $x$). }
The deformed differential calculus has more `constants' than the
ordinary one. From the above formulas one finds that $\d f=0$ if and
only if $f(t+\tau,x) = f(t,x)$ and $f(t,x+\lambda) = f(t,x)$. Let us
choose a point, say $0$, in $\Rl$ and construct the ideal $\I_\tau$ of
$\A$ generated by $f_\tau(t,x)-f_\tau(0,x)$ for all functions $f_\tau
\in \A$ which are periodic (with period $\tau$) in the first argument.
Then the algebra $\A/\I_\tau$ is isomorphic to
the algebra of real-valued functions on $\Ir \times \Rl$. If
$\I_{\tau,\lambda}$ denotes the ideal of functions generated by
$f_{\tau,\lambda}(t,x)-f_{\tau,\lambda}(0,0)$ for all functions periodic
in both arguments, one finds that $\A/\I_{\tau,\lambda}$ is isomorphic
to the algebra of real-valued functions on $\Ir^2$.

In what follows, depending on whether the parameters $\tau$ and
$\lambda$ are zero or not, $M$ denotes either $\Rl^2$, $\Rl \times \Ir$,
$\Ir \times \Rl$, or $\Ir^2$. Correspondingly, let $\A$
denote the algebra of real functions $f(t,x)$ which are smooth in both
arguments,  or functions $f_k(t) := f(t,k\lambda)$ smooth in the first
argument, respectively functions $f_k(x) := f(k \tau,x)$ smooth in
the second argument, or the algebra of all real functions $f_k(n) :=
f(n \tau , k \lambda)$ on $\Ir^2$. The differential calculus defined
above is then a differential calculus on the algebra $\A$.
Acting with the exterior derivative on (\ref{cr}) we obtain the
2-form relations
\be
     \d x  \, \d x = 0 \; , \qquad \d t \, \d x +\d x  \, \d t = 0 \; ,
     \qquad      \d t \, \d t = 0 \; .
\ee
Hence $\d t  \, \d x$ is a basis of the space of 2-forms $\O^2$ as a
left or right $\A$-module. There are no forms of higher grade, i.e.,
$\O^k=\{0\}$ for $k \geq 3$. Let $\O := \bigoplus_{k \geq 0} \O^k$
denote the differential algebra (where $\O^0 = \A$).

We introduce an inner product $( \; , \, ) \, : \, \O \times \O
\to \A$ via $(f,g) = f \, g$ for $f , g \in \A$,
\be
      (\d t , \d t) = -1 \; , \qquad (\d x , \d x) = 1 \; , \qquad
      (\d t , \d x) = 0 \; , \qquad
      (\d t \, \d x , \d t \, \d x) = - 1
\ee
and
\be
     (\psi, \phi) = (\phi,\psi) \; , \qquad
     (\psi, f \, \phi) = f \, (\psi, \phi)
\ee
for $\psi, \phi \in \O $.\footnote{Our inner product corresponds to
a metric with Lorentzian signature. The formalism works
as well with a Euclidean metric.}
If $\psi$ and $\phi$ have different grades,
then $(\psi, \phi)$ is set to zero.
As a consequence of these definitions we find
\be
      (\psi f , \phi) = (\psi , \phi f)  \; .
\ee
A Hodge $\ast$-operator can now be introduced as an $\Rl$-linear
operator on $\O$ via
\be
      (\d t \, \d x , \phi \ast \psi) := - (\phi,\psi) \; .
\ee
It satisfies the relations
\be
     \ast \, (\psi \, f) = f \, \ast \psi \; , \qquad
     \psi \ast \phi = \phi \ast \psi \; ,      \label{star-rel}
\ee
where $\phi$ and $\psi$ must have the same grade, and
\be
      \ast 1= \d t  \, \d x \; , \qquad
      \ast \, \d t = - \d x \; , \qquad
      \ast \, \d x = - \d t \; , \qquad
      \ast \, (\d t  \, \d x) = -1  \; .
\ee
Furthermore,
\be            \label{star-star}
    \ast \ast \, \psi(t,x) = (-1)^{r+1} \, \psi(t-\tau,x-\lambda)
\ee
for $\psi \in \O^r$.

The notion of an integral generalizes to our deformed differential
calculus in a natural way \cite{DMHS93}. In the following sections we
only need to consider one-dimensional integrals. It is therefore
sufficient here to define the integral for functions on $\Ir$.  An
indefinite integral is indeed determined by
\be
       \d \, \int^x f(x') \, \d x' = f(x) \, \d  x
\ee
up to a `constant', i.e., a periodic function.
A corresponding definite integral is then only defined over unions of
intervals which are integer multiples of $\lambda$. One obtains
\be          \label{integral}
      \int^{n \lambda}_{m \lambda}  f(x) \, \d x = \lambda \,
      \sum_{k=m}^{n-1} \, f(k \lambda)
\ee
 where $m,n \in \Ir$, $n > m$ (cf \cite{DMHS93}).

\section{Chiral models and conservation laws in two dimensions}
In this section we essentially follow Brezin et al \cite{Bre79}.
However, the following not only works for the ordinary differential
calculus but also for its deformations considered in the previous
section. $\B$ denotes a finite dimensional algebra of matrices and
$\B^\ast$ the group of invertible elements of $\B$. Let $g : M \to \B$
be invertible, i.e., $g(t,x) \in \B^\ast$ for all $t,x$. In terms of
\be
     A := g^{-1} \, \d g  \; ,
\ee
the field equations of a chiral model (generalized $\sigma$-model) are
\be
       \d \ast A = 0  \; .       \label{feqs}
\ee
The conservation laws of such a model are obtained inductively as
follows. Let $\Gamma$ be the space of fields $\Psi : M \to \B$ and
$D : \Gamma \to \O^1 \otimes_{\A} \Gamma$ the exterior covariant
derivative given by
\be
       D \Psi = \d \Psi + A \, \Psi  \; .
\ee
Since $A$ is a `pure gauge' we have
\be
       F := \d A + A A = 0  \; .
\ee
Moreover, using (\ref{feqs}) and (\ref{star-rel}) we find
\be
       \d \ast (A^i{}_j \Psi^j) = \d (\Psi^j \ast A^i{}_j)
    = (\d \Psi^j) \ast A^i{}_j = A^i{}_j \ast \d \Psi^j
\ee
and thus
\be               \label{dD}
      \d \ast D \Psi = D \ast \d \Psi \; .
\ee
Let $J^{(m)} : M \to \O^1 \oc \B$ be a conserved current, i.e.,
\be
       \d \ast J^{(m)} = 0   \; .
\ee
Since the first cohomology group of $M$ is trivial, i.e.,
$H^1(M)=\{0\}$, there exists $\chi^{(m)} : M \to \B$ such that
\be
         J^{(m)} = \ast \, \d \chi^{(m)} \; .         \label{dc}
\ee
Then
\be
         J^{(m+1)} := D \chi^{(m)}             \label{Dc}
\ee
is also conserved since
\be
    \d \ast J^{(m+1)} = \d \ast D \chi^{(m)}
    =  D \ast \d \chi^{(m)}
    =  D J^{(m)} = DD \chi^{(m-1)} = F \, \chi^{(m-1)} = 0   \; .
\ee
The induction starts with $\chi^{(0)} = I$, the unit matrix.
We then obtain an infinite number of conserved charges given by
\be
      Q^{(m)} := \int_{t=\cst} \ast J^{(m)} \;.
\ee
Let us calculate the first two of them. We have $J^{(1)}=DI=A$ and
therefore
\be         \label{Q1}
      Q^{(1)} = \int_{t=\cst} \ast A = - \int A_0(t-\tau,x) \, \d x
\ee
where $A=A_0 \, \d t + A_1 \, \d x$. Since $\ast \d \chi^{(1)} = J^{(1)}
=A$ we find $\d \chi^{(1)} = \ast A(t+\tau,x+\lambda)$ by use of
(\ref{star-star}). Hence
\be
      \chi^{(1)}(t,x) = -\int^x A_0(t,x'+\lambda) \, \d x'   \; .
\ee
and
\be
   \ast J^{(2)} = \ast D\chi^{(1)}
                = J^{(1)} + \ast \, (A \, \chi^{(1)}) \; .
\ee
Using (\ref{star-rel}) we find
\be
      Q^{(2)} = \int_{t=\cst} \ast J^{(2)} = \int A_1(t,x) \, \d x
  - \int A_0(t-\tau,x) \, \chi^{(1)}(t,x)  \, \d x  \; .
\ee

\vskip.2cm
         Let us introduce
\be
            \chi := \sum_{m=0}^\infty \gamma^m \, \chi^{(m)}
\ee
where $\gamma$ is a parameter. From (\ref{dc}) and (\ref{Dc}) we
obtain
\be
            \ast \, \d \chi^{(m+1)} = D \chi^{(m)}  \; .
\ee
Multiplying by $\gamma^{m+1}$ and summing over $m$ leads to
\be
            \ast \, \d \chi =\gamma \, D \chi   \; .  \label{in}
\ee
The field equations (\ref{feqs}) are integrability conditions of the
linear system (\ref{in}). This is seen as follows. Acting with $D$ on
(\ref{in}) and using $F=0$, we find $D \ast \d \chi=0$. Applying $\ast$
to (\ref{in}) we get $\d \chi(t-\tau,x-\lambda) = \gamma \ast D \chi$
and thus $\d \ast D \chi=0$ which, together with $D \ast \d \chi = 0$,
implies $\d \ast A=0$. Introducing $J := \ast \, \d \chi$ we
have
\be
      Q(t) := \int_{t=\cst} \ast J =
                \int_{t=\cst} \d \chi(t-\tau,x-\lambda)=
                \chi(t-\tau,x)|^{+\infty}_{-\infty}    \;  .
\ee

\section{Examples}
{\bf 1. Toda lattice.} Let $\B$ be the algebra $\Rl$ of real numbers
and let us write $g = e^{-q}$ with a function $q : M \to \Rl$.
The field equations (\ref{feqs}) then read
\be
               \d \ast (e^q \, \d e^{-q})  = 0  \; .      \label{Toda}
\ee
(a) $\tau=\lambda=0$. Then (\ref{Toda})  is just the wave equation $\d
\ast \d q = 0$, respectively,
\be
             \pa_t^2 q - \pa_x^2 q = 0   \; .
 \ee
\vskip.1cm
\noindent
(b) $\tau=0$ and $\lambda \neq 0$.  Then
\be
   A = - \dot{q}_k \, \d t + {1\over\lambda} \, (e^{q_k-q_{k+1}}-1)
               \, \d x   \, , \qquad
   \ast A = \dot{q}_k \, \d x - {1 \over \lambda} \, (e^{q_{k-1}-q_k}
                  -1) \, \d t
\ee
where $\dot{q}_k := dq_k/dt$. The field equations become
\be
     \ddot{q}_k = {1 \over \lambda^2} \left [ e^{q_{k-1}-q_k}-e^{q_k-
                        q_{k+1}} \right ]
\ee
which are those of the nonlinear Toda lattice \cite{Toda}. The conserved
charges can be obtained using the method described in the previous
section.
In particular, evaluation of (\ref{Q1}) using (\ref{integral}) leads to
\be
           Q^{(1)} = \lambda \sum_{k=-\infty}^\infty \dot{q}_k
\ee
which is the total momentum. Furthermore,
\be
     \chi^{(1)}(t,k \lambda) = \lambda \, \sum_{\ell=-\infty}^k
     \dot{q}_\ell(t)
\ee
(modulo addition of a constant) and thus
\be
     Q^{(2)} = \sum_{k=-\infty}^\infty \left [ e^{q_k-q_{k+1}}-
    1+\lambda^2 \sum_{\ell \leq k} \dot{q}_k\dot{q}_\ell \right]  \; .
\ee
This yields
\be
     Q^{(2)}- {1\over2} [Q^{(1)}]^2 =
     \sum_k \, [ {1\over2} \lambda^2 \dot{q}_k^2 + e^{q_k-q_{k+1}} - 1 ]
\ee
which is the total energy.
\vskip.1cm
\noindent
(c) $\tau \neq 0, \, \lambda \neq 0$. Then
\be
      A &=& {1\over\tau} (e^{q_k(n)-q_k(n+1)}-1) \, \d t
             + {1\over\lambda} (e^{q_k(n)-q_{k+1}(n)}-1) \, \d x
                      \label{Toda_A}    \\
     \ast A &=& -{1 \over \tau} (e^{q_k(n-1)-q_k(n)}-1) \, \d x
                  -{1 \over \lambda} (e^{q_{k-1}(n)-q_k(n)}-1) \, \d t
\ee
and the field equations are
\be
    {1 \over \tau^2} \, \left [ e^{q_k(n-1)-q_k(n)}-e^{q_k(n)-
     q_k(n+1)} \right ]
   = {1 \over \lambda^2} \, \left[ e^{q_{k-1}(n)-q_k(n)}
     - e^{q_k(n)-q_{k+1}(n)} \right ]    \; .
\ee
This describes a discrete-time Toda lattice.\footnote{See also
\cite{Suri90} and references given there for a class of discrete time
generalized Toda lattices.}
Written in the form
\be
     e^{q_k(n) - q_k(n+1)} = e^{q_k(n-1)-q_k(n)}
    - {1 \over c^2} \, \left[ e^{q_{k-1}(n)-q_k(n)}
    - e^{q_k(n)-q_{k+1}(n)} \right ]
\ee
with $c := \lambda/\tau$,
we see that the rhs is not necessarily positive in contrast to the lhs.
As a consequence, there is a constraint on initial values.
Exact solutions are given by
\be                 \label{Toda_left_movers}
      q_k(n) = f(k+n) - 2 \, k \, \mbox{ln} \, c
\ee
where $f$ is an arbitrary function. These `left movers' do not have a
right-moving counterpart.
{}From (\ref{Toda_A}) and (\ref{Dc}) it is evident that $q_k(n)$
enters the conserved charges only through the quantities
\be
      U_k(n) := e^{q_k(n) - q_{k+1}(n)}  \, , \qquad
      V_k(n) := e^{q_k(n) - q_k(n+1)}
\ee
for which we obtain the following first order system,
\be
     U_k(n+1) &=& { V_{k+1}(n) \over V_k(n) } \, U_k(n)  \, ,
                                      \nonumber \\
     V_k(n+1) &=& V_k(n) + c^{-2} \, \lbrack  U_k(n+1) - U_{k-1}(n+1)
                            \rbrack  \; .
\ee
The first conserved charge is
\be
       Q^{(1)}(n) = c \, \sum_k \lbrack 1 - V_k(n) \rbrack   \; .
\ee
When $c \neq 1$, the solutions (\ref{Toda_left_movers}) become
infinite either for $k \to \pm \infty$ or for $n \to \pm \infty$.
The quantities $U_k(n), V_k(n)$ and thus also the conserved charges
may remain finite, however.

\vskip.2cm
\noindent
{\bf 2. $GL(n,\Rl)$-models.}  We express an element $g \in GL(n,\Rl)$ as
$g = \pm e^{-q} \, s$ with real $q$ and $s \in SL(n,\Rl)$. In the
following we only discuss the case $\tau = 0$ and $\lambda \neq 0$.
The field equations (\ref{feqs}) then split into the two parts,
\be
   \ddot{q}_k = {1\over n \, \lambda^2} \, \left \lbrack
   e^{q_{k-1}-q_k} \, \mbox{tr}(s^{-1}_{k-1}s_k)
    - e^{q_k-q_{k+1}} \, \mbox{tr}(s^{-1}_k s_{k+1}) \right \rbrack
\ee
and
\be
     (s^{-1}_k \dot{s}_k)\, \dot{}  = {1\over \lambda^2} \left (
     e^{q_k-q_{k+1}} \, [ s^{-1}_k \, s_{k+1} - {I \over n} \,
     \mbox{tr}(s^{-1}_k s_{k+1}) ]
    - e^{q_{k-1}-q_k} \, [ s^{-1}_{k-1}s_k - {I \over n} \,
    \mbox{tr}(s^{-1}_{k-1}s_k) ]  \right )
\ee
(where $I$ is the unit matrix).
The first equation resembles that of the nonlinear Toda lattice to which
it reduces for $s$ not depending on $k$.

\vskip.2cm
\noindent
{\bf 3.  A $GL(1,\Cx)$-model.}  We write an element of $GL(1,\Cx)$ in
the form $g = e^{-q} \, e^{i \theta}$ with real $q$ and $\theta$. In
this case the field equations for $\tau = 0$ and $\lambda \neq 0$ read
\be
  \ddot{q}_k &=& {1\over \lambda^2} \left \lbrack  e^{q_{k-1} - q_k} \,
  \cos(\theta_k-\theta_{k-1}) - e^{q_k-q_{k+1}} \,
  \cos(\theta_{k+1}-\theta_k)  \right \rbrack \, ,     \\
  \ddot{\theta}_k &=& {1\over \lambda^2} \left \lbrack
  e^{q_{k-1}-q_k} \, \sin(\theta_k - \theta_{k-1})
  - e^{q_k-q_{k+1}} \, \sin(\theta_{k+1} - \theta_k) \right \rbrack \; .
\ee
A reduction to a $U(1)$-model by setting $q = 0$ leads to a constraint.
On the other hand, setting $\theta = 0$ simply leads us back to the
nonlinear Toda lattice.

\vskip.2cm
\noindent
{\bf 4.  A $GL(1,\Ham)$-model.} Here $\Ham$ denotes the quaternions.
Again, we write an element in the form $g = e^{-q} \, u$ with real $q$
and $u = \alpha + \beta \, i + \gamma \, j + \delta \, k$, $u \bar{u} =
\alpha^2 + \beta^2 + \gamma^2 + \delta^2 = 1$. For $\tau = 0$ and
$\lambda \neq 0$ the field equations (\ref{feqs}) take the form
\be
  \ddot{q}_k &=& {1 \over 2 \, \lambda^2} \left \lbrack e^{q_{k-1}-q_k}
  \,  ( \bar{u}_{k-1} u_k + \bar{u}_k u_{k-1} )
  - e^{q_k-q_{k+1}} \,  ( \bar{u}_k u_{k+1} + \bar{u}_{k+1}  u_k )
  \right \rbrack   \, , \\
  (\bar{u}_k \dot{u}_k) \, \dot{} &=& {1 \over 2 \, \lambda^2}
  \left \lbrack e^{q_{k-1}-q_k} \,  (\bar{u}_{k-1} u_k
  - \bar{u}_k u_{k-1} ) - e^{q_k-q_{k+1}} \,
  (\bar{u}_k u_{k+1} - \bar{u}_{k+1} u_k) \right \rbrack   \; .
\ee
Although $GL(1, \Ham)$ is isomorphic with $SL(1, \Rl) \times U(2)$, if
we had chosen the latter representation, some unpleasant constraints
would have shown up.

\vskip.2cm
\noindent
{\bf 5. The $O(n)$ $\sigma$-model.} Let $\S \in \Rl^n$ be a unit vector,
$\S \bu \S=1$.  The matrix $g = I-2P$ with $P=\S \oc \S$ is orthogonal.
We have $P^2=P$ and therefore $g^{-1}=g$. Hence
\be
      A = 2 \, (P \, \d P - \d P \, P)
\ee
using $\d P = \d P^2 = P \, \d P + \d P \, P$. In terms of $\S$ we have
\be
 A = 2 \, [\S \oc \d \S - \d \S \oc \S + 2 \, \S \oc (\S \bu \d \S) \S ]
\ee
where we used $\d \S \bu \S + \S \bu \d \S = 0$.\footnote{Note that
(for $\tau \neq 0$ or $\lambda \neq 0$) the noncommutativity between
functions and differentials leads to $\S \bu \d \S \neq 0$, in general.}
For $\tau \neq 0$ and $\lambda \neq 0$, we obtain
\be
   \ast A &=& - 2 \lbrack \S^{-x} \oc \pa_{-x} \S - \pa_{-x} \S \oc \S
   + 2 \, (\S^{-x} \bu \pa_{-x} \S) \, \S^{-x} \oc \S  \rbrack \, \d t
                        \nonumber   \\
   & &  - 2 \lbrack \S^{-t} \oc \pa_{-t} \S - \pa_{-t} \S \oc \S
   + 2 \, (\S^{-t} \bu \pa_{-t} \S) \, \S^{-t} \oc \S  \rbrack \, \d x
\ee
with $\S^{-t}(t,x) := \S(t-\tau,x)$ and a corresponding definition for
$\S^{-x}$. The field equations $\d \ast A = 0$ now take the form
\be
    \pa_{+t} \left [ \S^{-t} \oc \pa_{-t} \S - \pa_{-t} \S \oc \S +
  2 \, (\S^{-t} \bu \pa_{-t} \S) \, \S^{-t} \oc \S \right ] - & &
              \nonumber \\
  \pa_{+x} \left [ \S^{-x} \oc \pa_{-x} \S - \pa_{-x} \S \oc \S +
  2 \, (\S^{-x} \bu \pa_{-x} \S) \, \S^{-x} \oc \S \right ] &=& 0 \; .
\ee
In the limit $\tau, \lambda \to 0$ we get
\be
     \S \oc (\pa_t^2 - \pa_x^2) \S = (\pa_t^2 - \pa_x^2) \S \oc \S  \; .
\ee
Acting on this equation from the left with $\S \bu$ and using
$\S \bu \d \S = 0$ (which holds in the case $\tau=\lambda=0$), we
recover the field equations of the classical nonlinear $\sigma$-model on
$\Rl^2$,
\be
     \pa_t^2 \S - \pa_x^2 \S + (\pa_t \S \bu \pa_t \S) \S
    - (\pa_x \S \bu \pa_x \S) \S = 0    \; .
\ee
The conserved charges of this model were first obtained in \cite{LP78}
by means of inverse scattering methods.
When $\tau = 0$ and $\lambda \neq 0$, the field equations are
\be
          \S_k \oc \pa_t^2 \S_k - \pa_t^2 \S_k \oc \S_k
  &=& {1 \over \lambda^2} \, \lbrack 2 \, (-\S_k \bu \S_{k+1} ) \,
          \S_k \oc \S_{k+1} - 2 \, ( \S_{k-1} \bu \S_k) \, \S_{k-1}
          \oc \S_k          \nonumber \\
   & & - \S_{k+1} \oc \S_{k+1} + \S_{k-1} \oc \S_{k-1} \rbrack
             \label{latt-sigma}
\ee
in terms of $\S_k(t) := \S(t,\lambda k)$.
Acting from the left\footnote{Acting from the right with $\bu \S_k$
yields the same equation but with $\S_{k+1}$ replaced by $\S_{k-1}$
in the quadratic term on the rhs.}
with $\S_k \bu$ and using $\S_k \bu \S_k = 1$
(which implies $\S_k \bu \pa_t^2 \S_k = - \pa_t \S_k \bu \pa_t \S_k$)
we obtain
\be
    \ddot{\S}_k + (\dot{\S}_k \bu \dot{\S}_k) \S_k +
  {1 \over \lambda^2} \left [ 2 \, (\S_k \bu \S_{k-1})^2 \, \S_k -
  (\S_k \bu \S_{k+1}) \S_{k+1} - (\S_k \bu \S_{k-1}) \S_{k-1}
  \right ] = 0  \; .              \label{S-evolution}
\ee
This does not exhaust the equations (\ref{latt-sigma}), however. In
addition we have the constraint
\be
      \S_{k+1} \oc \S_{k+1} - \S_{k-1} \oc \S_{k-1}
     + (\S_{k-1} \bu \S_k) \, \S_k \oc \S_{k-1}
     - (\S_k \bu \S_{k+1}) \, \S_k \oc \S_{k+1}  & &   \nonumber \\
     + (\S_{k-1} \bu \S_k) \, \S_{k-1} \oc \S_k
     - (\S_k \bu \S_{k+1}) \, \S_{k+1} \oc \S_k &=& 0  \, . \qquad
\ee
The problem is that this constraint is not automatically respected by
(\ref{S-evolution}) so that differenti\-ation with respect to time
generates additional equations of motion. The conclusion is that the
equations governing our lattice $O(n)$ $\sigma$-model are not `good'
equations. There are at least simple exact solutions like those given by
\be
      \S_k \bu \S_{k+1} = 0 \, , \quad
      \S_{k+2} = \pm \S_k \, , \quad
       \ddot{\S}_k + ( \dot{\S}_k \bu \dot{\S}_k ) \, \S_k = 0 \; .
\ee
\vskip.2cm

The appearance of a constraint, as in our last example, is a rather
general feature which can be understood as follows. The equation
$\d \ast (g^{-1} \d g) = 0$, where $g$ has values in a group, takes
the following form for $\tau = 0$ and $\lambda \neq 0$,
\be
     \pa_t ( g_k^{-1} \, \pa_t g_k ) = {1 \over \lambda^2} \, ( g^{-1}_k
     \, g_{k+1} - g^{-1}_{k-1} \,  g_k ) \; .
\ee
Whereas the lhs is in the Lie algebra of the group, the rhs lives in
the group algebra. The deviation of the group algebra from the Lie
algebra results in constraints.
In case of the orthogonal group $O(n)$, the lhs of the last equation
is an antisymmetric matrix, but this property is not shared by
the rhs. The symmetric part of the matrix equation is then a
constraint. This problem does not appear for $GL(n)$-models.
In case of the $O(n)$-model, one may take into consideration a
simultaneous deformation of $\S \bu \S = 1$ as an
attempt to get rid of the constraints.

\section{Lax pairs for the chiral models}
In this section we construct Lax pairs for our chiral models.
This establishes contact with other formulations of integrable models
(see \cite{Fadd+Takh87}, for example). The starting point is the
equation $\ast \, \d \chi = \gamma \, D \chi$ derived in section 3.
\vskip.2cm
\noindent
{\bf 1) $M = \Rl^2$.} Evaluation of (\ref{in}) leads to
\be
   \pa_t \chi = -\gamma \, [\pa_x \chi + (g^{-1} \pa_x g) \, \chi ] \; ,
\qquad
   \pa_x \chi = -\gamma \, [\pa_t \chi + (g^{-1} \pa_t g) \, \chi ] \; .
\ee
Solving this system for the partial derivatives of $\chi$, we obtain
\be          \label{linsyst_cont}
    \pa_x \chi = L \, \chi \; ,   \qquad    \pa_t \chi = M \, \chi
\ee
with
\be
 L(t,x;\gamma) = {\gamma \over1 - \gamma^2} \, g^{-1} \, (-\pa_x
     g + \gamma\, \pa_t g) \; ,   \qquad
 M(t,x;\gamma) = {\gamma \over1 - \gamma^2} \,
 g^{-1} \, (\gamma \, \pa_x g - \pa_t g)   \; .
\ee
In terms of $L$ and $M$ the integrability conditions for the system
(\ref{linsyst_cont}), which are the field equations, read
\be
     \pa_t L - \pa_x M + \lbrack L , M \rbrack = 0  \; .
\ee
\vskip.2cm
\noindent
{\bf 2) $M = \Rl\times\Ir$.} In this case (\ref{in}) yields
\be
 \dot{\chi}_k = {\gamma \over \lambda} \, [ g^{-1}_k \,
     g_{k+1} \, \chi_{k+1} - \chi_k ] \; ,  \qquad
 \chi_k - \chi_{k-1} =  -\gamma \lambda \, [ \dot{\chi}_k + g^{-1}_k \,
 \dot{g}_k \, \chi_k ] \; .
\ee
Introducing $\psi_k := \chi_{k-1}$, $\phi_k := \gamma \, g_k \, \chi_k$
and $\xi_k := (\phi_k , \psi_k)^T$, we obtain
\be               \label{linsyst_cd}
    \xi_{k+1} = L_k \, \xi_k \; ,  \qquad \dot{\xi}_k = M_k \, \xi_k
\ee
with
\be
    L_k(t;\gamma) =  \gamma^{-1} \, \left ( \begin{array}{cc}
    (\gamma^{-1} + \gamma) \, I + \lambda \, \dot{g}_k \,
    g^{-1}_k   & - g_k \\  g^{-1}_k  & 0 \end{array} \right )
\ee
and
\be
 M_k(t;\gamma) = \lambda^{-1} \,  \left ( \begin{array}{cc}
    - \gamma^{-1} \, I & g_k \\
    -  g^{-1}_{k-1}  &  \gamma \, I    \end{array} \right) \; .
\ee
The field equations of the chiral model are now obtained as integrability
conditions of the linear system (\ref{linsyst_cd}) in the form
\be
     \pa_t L + L_k \, M_k - M_{k+1} \, L_k = 0  \; .
\ee
\vskip.2cm
\noindent
{\bf 3) $M = \Ir^2$.} Now (\ref{in}) leads to
\be
  \chi_k(n) - \chi_k(n-1) & = & -{\gamma \over c} \,
  [ g_k(n)^{-1} \,  g_{k+1}(n) \, \chi_{k+1}(n) - \chi_k(n) ]   \, ,
                                                          \nonumber  \\
  \chi_k(n) -\chi_{k-1}(n) & = & -\gamma c \,
  [ g_k(n)^{-1} \, g_k(n+1) \, \chi_k(n+1) - \chi_k(n) ]
\ee
where $c:=\lambda/\tau$. Let us introduce $\psi_k(n) := \chi_{k-1}(n-
1)$, $\phi_k(n) := \gamma \, g_k(n) \, \chi_k(n)$ and
$\xi_k(n) := (\phi_k(n),\psi_k(n))^T$. Then
\be
      \xi_{k+1}(n) = L_k(n) \, \xi_k(n)  \; , \qquad
      \xi_k(n+1) = M_k(n) \, \xi_k(n)
\ee
with
\be
  L_k(n;\gamma) = (1 - \gamma c )^{-1} \, \hat{L}_k(n;\beta)  \, ,
      \qquad
  M_k(n;\gamma) =  (c - \gamma )^{-1} \,  c^{-1} \, \hat{M}_k(n;\beta)
\ee
where
\be
      \hat{L}_k(n;\beta) &=& \left ( \begin{array}{cc}
     \beta \, I - c^2 \, g_k(n) \, [g_k(n-1)]^{-1}  &  c \, g_k(n)  \\
    - c \, [g_k(n-1)]^{-1}  &  I            \end{array} \right )
                          \nonumber  \\
 \hat{M}_k(n;\beta) &=&  \left (       \begin{array}{cc}
     \beta \, I - g_k(n) \,  [g_{k-1}(n)]^{-1}  &  c \, g_k(n) \\
     - c \, [g_{k-1}(n)]^{-1}  &  c^2 \, I  \end{array} \right )
           \label{LMhat}
\ee
and $\beta = 1 - c \, (\gamma + \gamma^{-1}) + c^2$.
Now the field equations are obtained as
\be                         \label{feqs_hatLM}
  \hat{L}_k(n+1) \, \hat{M}_k(n) = \hat{M}_{k+1}(n) \, \hat{L}_k(n)
\ee
(suppressing the argument $\beta$). The matrices $\hat{L}_k$ and
$\hat{M}_k$ are invertible if $\beta \neq 0$.

\section{B\"acklund transformations for the chiral models}
If $g$ is an exact solution of the field equation for a chiral model,
then also $h$ if
\be            \label{Baecklund_F}
      h^{-1} \d h - g^{-1} \d g =  \ast \, \d F
\ee
with a $\B$-valued function $F$. A suitable choice for $F$ is essential
for this relation to be useful for generating new solutions from given
solutions. The usual continuum ($M = \Rl^2$) B\"acklund transformation
is obtained for $F = \beta \, h^{-1} g$ where $\beta$ is the spectral
parameter (see \cite{OPS80}, for example).
For the nonlinear Toda lattice the relation
\be              \label{Toda_Baecklund}
      h^{-1}_k \, \d h_k -  g^{-1}_k \,  \d g_k =  \beta \, \ast \, \d
      (h_k^{-1} \, g_{k+1})
\ee
reproduces the corresponding formulas in \cite{Toda}. After some
manipulations we find, for $\tau = 0$ and $\lambda \neq 0$,
\be
  g^{-1}_k \dot{g}_k & = & {1 \over \lambda} \left ( \beta \,
  h^{-1}_{k-1} g_k + {1\over\beta} \, g^{-1}_k h_k +\alpha \, I
  \right )  \, ,                      \nonumber   \\
  h^{-1}_k \dot{h}_k & = & {1 \over \lambda} \left ( \beta \,  h^{-1}_k
  g_{k+1} + {1\over\beta} \, g^{-1}_k h_k +\alpha \, I \right )
\ee
where $\alpha$ is a constant of integration. Surprisingly, written in
this way it generalizes to a B\"acklund transformation for the general
chiral model on $M = \Rl \times \Ir$. However, for a
noncommutative group $\B^\ast$ the above set of equations is no
longer equivalent to (\ref{Toda_Baecklund}),  though there must be
some $F$  so that (\ref{Baecklund_F}) is satisfied.
\vskip.2cm

Now we turn to the remaining case $M = \Ir^2$. Here we proceed in a
different way.  For an invertible matrix
\be
 {\cal M} = \left( \begin{array}{cc} a & b \\ c & d \end{array} \right)
\ee
with entries $a,b,c,d \in \B$ and for $z \in \B$ we define
${\cal P}({\cal M}) \, z := (a z + b) (c z + d)^{-1}$.
Using ${\cal P} ({\cal M}) {\cal P} ({\cal M}') = {\cal P} ({\cal
M} {\cal M}')$ the field equations in the form (\ref{feqs_hatLM}) can
now be rewritten as
\be
   {\cal P} (\hat{L}_k(n+1)) \, {\cal P} (\hat{M}_k(n)) = {\cal P}
   (\hat{M}_{k+1}(n)) \, {\cal P} (\hat{L}_k(n))   \; .
\ee
For $h_k(n) \in \B^\ast$ we define
\be
     h_{k+1}(n) = {\cal P} (\hat{L}_k(n)) \, h_k(n) \, , \qquad
     h_k(n+1)  = {\cal P} (\hat{M}_k(n)) \, h_k(n)  \; .
\ee
The field equations for $g_k(n)$ are now recovered as integrability
conditions of this system, i.e., by calculating $h_{k+1}(n+1)$ from
both of the last two equations. The latter can be written in the form
\be
    h_{k+1}(n) & = & c \, g_k(n) + \beta \, ( c \, [g_k(n-1)]^{-1} -
    [h_k(n)]^{-1} )^{-1}   \nonumber  \\
    c \, h_k(n+1) & = &  g_k(n) +
    \beta \, ( [g_{k-1}(n)]^{-1} -  c \, [h_k(n)]^{-1} )^{-1}
                                \label{Toda_dt_BT}
\ee
or, equivalently,
\be
   g_{k+1}(n)  & = & c \, h_{k+1}(n+1) + \beta \, ( c \,
   [h_{k+1}(n)]^{-1} - [g_k(n)]^{-1} )^{-1}   \nonumber \\
   c \, g_k(n+1)  & = &  h_{k+1}(n+1) +
  \beta \, ( [h_k(n+1)]^{-1} - c \, [g_k(n)]^{-1} )^{-1}    \; .
\ee
The last set of equations shows that the field equations for $g_k(n)$
are also obtained by calculating $g_{k+1}(n+1)$ from both equations and
comparing the results. As a consequence, if $g_k(n)$ is a solution,
then also $h_k(n)$. Hence (\ref{Toda_dt_BT})  defines
a B\"acklund transformation. Applied to the left movers
(\ref{Toda_left_movers}) of the Toda model on $\Ir^2$ and writing
$h_k(n) = e^{-q'_k(n)}$, the transformation leads to
$q'_k(n+1) = c^{-2} \, q'_{k+1}(n)$ (together with a constraint on
initial values) and thus again to left movers.

\section{Conclusions}
We have presented a general method for integrable discretizations of
two-dimensional chiral models via deformations of
the ordinary differential calculus. As an example, the nonlinear Toda
lattice is obtained in this way from the linear wave equation. It may be
regarded as a $GL(1,\Rl)$ chiral model on $\Rl \times \Ir$. More
generally, our method works well in particular for $GL(n,\Rl)$ and
$GL(n,\Cx)$ models. In case of a chiral model for which the group
algebra differs from the Lie algebra, the resulting discretized model
in general suffers from unpleasant constraints, however. This has been
demonstrated for the $O(n)$ $\sigma$-model. Perhaps there are
modifications of our formalism which can improve such models.
\vskip.1cm

The general formalism presented here also works for
corresponding deformations of the differential calculus on $\Rl \times
S^1$ to calculi on a periodic (space) lattice. In that case, however,
the first cohomology group is no longer trivial and (\ref{dc}) must no
longer hold. But if the continuum chiral model is integrable, then also
its deformations.
Furthermore, one may consider other (and in particular
curved) metrics on the underlying two-dimensional space.
In fact, the possibilities to generalize the formalism developped in
this work extend much beyond what we have mentioned so far.
The deformed differential calculus of section 2 may be replaced by
other differential calculi with a two-dimensional space of 1-forms
(see \cite{DMH95} for candidates). All we need is a suitable
generalization of the $\ast$-operator. It is then possible to
generalize the (continuum) definition of a chiral model. This is what
we have done for a restricted class of differential calculi. It turned
out that the nonlinear Toda lattice belongs to the corresponding
extended class of chiral models. It remains to be seen whether
other integrable models can also be understood as generalized chiral
models.

\vskip.2cm
\noindent
{\bf Acknowledgment}. F M-H thanks M. Bordemann for a motivating
discussion.

\end{document}